# Magneto-optical resonances in fluorescence from sodium D$_2$ manifold

Raghwinder S. Grewal[1], Gour Pati[1], Renu Tripathi[1,*], Anthony W. Yu[2], Michael Krainak[2] and Michael Purucker[2]

[1]*Division of Physics, Engineering, Mathematics and Computer Science*
*Delaware State University, Dover, DE 19901, USA*
[2]*NASA GSFC, Greenbelt, MD 20771, USA*

We report on magneto-optical resonances observed in sodium fluorescence from D$_2$ manifold with an intensity modulated light. Fluorescence resonances are measured in the perpendicular (90º) and backward (180º) directions to the light propagation, in laboratory experiments using a sodium cell containing neon buffer gas. Properties of these resonances are studied by varying the magnetic field at fixed light modulation frequency, and vice-versa. Modulation with low-duty cycle shows higher-harmonic resonances of the modulation frequency and sub-harmonic resonances of the Larmor frequency. A dark resonance with maximum amplitude for laser wavelength closer to the crossover peak is observed. The origin of this dark resonance observed in Na D$_2$ line is discussed using a theoretical model. Present study is aimed towards improving the understanding of magneto-optical resonances for remote magnetometry applications with mesospheric sodium.

## I. INTRODUCTION

Return fluorescence obtained from laser excitation of mesospheric sodium is traditionally used as laser guide star to correct wavefront distortion in astronomical observations [1,2]. Recently, laser excitation of mesospheric sodium has received considerable attention in remote sensing of geomagnetic field [3-6]. The method is based on the measurement of spin precession of mesospheric sodium atoms by polarizing them through synchronous optical pumping with a modulated laser beam. Long-time measurement and large-scale mapping of geomagnetic field can provide better insight into various magnetic phenomena impacting the mesosphere region [7-9].

Coherent interaction of an atomic medium with an optical field creates light-induced atomic coherence through the coupling of different Zeeman sublevels or hyperfine levels of atoms in the medium. These atomic coherences can create various resonant effects in the medium [10-15]. Among these, dark and bright magneto-optical resonances have been widely studied [16-18]. Typically, a dark resonance is observed due to trapping of atoms in a dark state resulting from the coupling of two ground-state sublevels via a common excited-state sublevel. On the other hand, coupling among the excited-state sublevels results in the formation of a bright resonance via transfer of atomic coherence through spontaneous emission [18-20].Unlike other alkali atoms (K, Rb, Cs), magneto-optical resonance in sodium atom is not extensively studied. Understanding light-atom interaction and the origin of magneto-optical resonances in sodium via laboratory experiments is important for remote magnetometry sky experiments. Recently, magnetic resonance in sodium D$_1$ line using is studied in a sodium cell with high buffer gas pressure [21]. The authors showed a dark resonance with reduced absorption at the center of resonance.

In this paper, we investigate magneto-optical resonances in sodium D$_2$ line fluorescence, which has been widely used in sky experiments [4-6]. We study the properties of magnetic resonances by detecting fluorescence in directions perpendicular (90º) and backward (180º) to laser beam propagation in a sodium cell. Fluorescence measurement in backward direction resembles the sky experimental configuration. Resonances are acquired using two independent methods: phase-sensitive lock-in detection performed using a high bandwidth lock-in amplifier, and direct amplitude measurement performed using a low-pass filter (or integrator). We measured the resonances as functions of magnetic field and laser modulation frequency (or pulse repetition rate) with different pulse duty cycles. Our measurements show that low-pass filtering can provide a convenient method for detecting magnetic resonances by sweeping laser modulation frequency over a wide range. Instead of a bright resonance, a dark resonance is observed near the cyclic $F_g = 2 \rightarrow F_e = 3$ (here 'g' refers to the ground, and 'e' to the excited state). The maximum amplitude of dark resonance is observed near the ground-state crossover resonance. We explain the origin of this dark resonance using a theoretical density-matrix model including buffer gas induced collisional dephasing in the excited state. We have also studied the effect of repump beam to show that population transfer by repump can increase the amplitude of magnetic resonance.

*rtripathi@desu.edu

## II. EXPERIMENTAL SETUP

The schematic of the experimental setup is shown in Fig. 1. A narrow-band frequency-doubled Raman fiber amplifier laser tuned to sodium (Na) $D_2$ line is used in the experiment. A combination of half-wave plate and a polarizing beam splitter (PBS) is utilized to control the optical power in the laser-lock setup. Two different methods are employed to monitor the laser wavelength using buffer gas free Na reference cells. First, the Doppler-broadened fluorescence spectrum at 90° angle to light propagation direction is collected using an avalanche photodiode (APD). Using an offset lock with the help of a proportional integral (PI) controller (Newport LB1005 Servo Controller), the laser is locked on different positions in Na $D_2$ line by utilizing the Doppler-broadened fluorescence spectrum from the APD. Second, the Doppler-free absorption peaks/dips are observed in transmitted light from a saturation absorption spectroscopy (SAS) setup. This SAS setup is used only to find the exact position of the laser with respect to $F_g = 2 \rightarrow F_e (= 1, 2, 3)$ [referred as $D_{2a}$ transition], $F_g = 1 \rightarrow F_e (= 0, 1, 2)$ [referred as $D_{2b}$ transition] and the crossover peak present in the Doppler-free absorption spectrum [22].

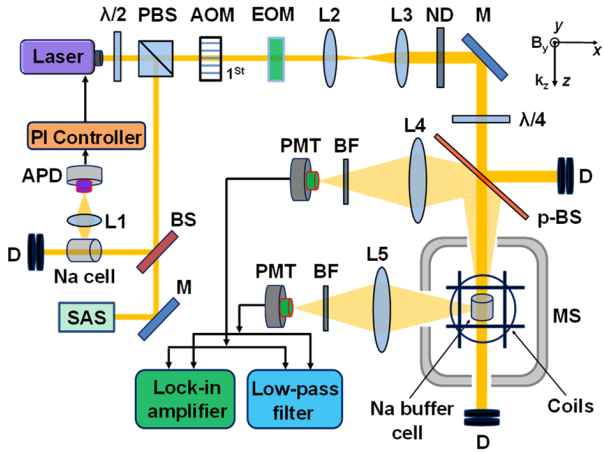

FIG. 1. Schematic diagram of the experimental setup. λ/2, half-wave plate; λ/4, quarter-wave plate; PBS, polarizer beam splitter; BS, beam splitter; D, beam dump; M, mirror; L1-L5, convex lenses; BF, band-pass filter; PMT, photomultiplier tube.

The laser beam intensity is modulated using an acoustic-optic modulator (AOM) driven by the rectangular pulse waveform with different duty cycles, and the first-order diffracted beam is selected as the probe beam for the experiment [23]. A resonant electro-optic phase modulator (EOM) shown in the setup, is only used while studying the effect of optical repump on the amplitude of magnetic resonance. This EOM is set to produce optical sidebands as repump at frequencies ±1710 MHz with respect to the carrier (or probe beam) frequency. The Na cell used in our fluorescence experiments contains 10 Torr neon (Ne) buffer gas. The laser beam diameter is expanded to approximately 8 mm (using two lenses L2 & L3) prior to passing through the Na cell, to increase the number of interacting Na atoms in the beam volume and prevent transit-time broadening of the magnetic resonance.

A neutral density (ND) filter is used to control the probe beam power to the cell. Using a λ/4 plate, the light polarization is set to circular. A non-polarizing pellicle beam splitter of diameter 2" transmits ~50% laser power onto the experimental Na buffer cell. The experimental cell is kept inside a two-layered magnetic shield (μ-metal) enclosure, which reduces the ambient magnetic field by a factor of ~$10^2$. Three pairs of Helmholtz coils installed inside the shield help us to further reduce the ambient field by canceling the residual magnetic field inside the enclosure. Coils along y-axis perpendicular to the light propagation vector ($k_z$) are utilized to apply fixed or scanning magnetic field, as per the experimental needs. Temperature of the Na buffer cell is kept at 86° Celsius to keep sodium atomic density at low levels (~$1.0 \times 10^9$ atoms/$cm^3$ [24]), thus making it comparable to the number of interacting Na atoms in the mesosphere column (length ≃ 10 km). Magnetic resonances are obtained by collecting fluorescence light from the Na buffer cell at two different angles. First, fluorescence at 90° angle to the light propagation direction is detected using a PMT at the focal plane of the lens L5. Second, fluorescence at 180° angle (i.e. backward) to the light propagation direction is detected using another PMT at the focus of lens L4. The backscattered fluorescence is separated from the oncoming laser beam by using the pellicle beam splitter (p-BS). To eliminate the background light from the fluorescence signal, an ultra-narrow band-pass filter (Alluxa 589.45-1 OD6) is used before the PMT.

The magnetic resonance signals are observed by two independent methods: (a) demodulating the detected signal with a lock-in amplifier (Stanford Research System SR865), operating at the first harmonic of the laser modulation frequency $\Omega_m$ and (b) measuring the detected signal directly through a low-pass filter (Stanford Research Systems SR560).

## III. EXPERIMENTAL RESULTS AND DISCUSSION

Fluorescence measurements are first performed in the direction perpendicular to light propagation and by scanning the magnetic field ($B_y$) along $y$-axis while keeping $\Omega_m$ fixed. Figure 2 shows magneto-optical resonances observed corresponding to three different duty cycles of light modulation. The probe beam is modulated at fixed $\Omega_m$ = 20 kHz frequency with an average intensity of 0.25 mW/cm$^2$. The EOM is kept off during these measurements, and the probe beam is locked closer to the crossover peak (observed midway between $D_{2a}$ and $D_{2b}$ transitions) using offset laser lock where maximum amplitude of the magneto-optical resonance is observed. Fluorescence signal is demodulated using the lock-in amplifier; the in-phase and the quadrature-phase components are shown respectively, in Fig. 2(a, first column) and 2(b, second column).

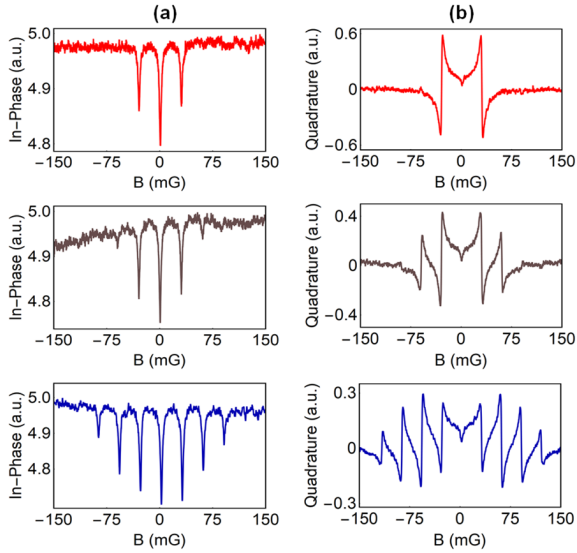

FIG. 2. Magneto-optical resonances: in-phase (a) and quadrature (b) components of the lock-in amplifier are measured at different modulation duty cycles; 50 % (top row), 35 % (middle row) and 20 % (bottom row) at $\Omega_m$ = 20 kHz.

When the laser is modulated with 50% duty cycle [Fig. 2 (top row)], synchronous optical pumping of atoms leads to optical resonances at Larmor frequency $\pm \Omega_L$ ($\gamma B_y = \Omega_m$) [25], as seen in both in-phase and quadrature-phase signals, where $\gamma$ is the Gyromagnetic ratio. The in-phase signal shows the amplitude, and the quadrature signal shows the phase of the resonance. The dips at resonant Larmor frequency indicate dark resonances in the in-phase signal. The width (i.e. full width at half maximum, FWHM) of the resonance near zero magnetic field is measured to be approximately 3.2 mG (2.24 kHz). As the duty cycle of modulation is lowered to 35% [Fig. 2 (middle row)], the second harmonic signals corresponding to $\Omega_L = \pm 2\Omega_m$ are also seen along with the first harmonic. These higher-order resonances at lower duty cycles are formed due to the presence of Fourier components at integer multiples of $\Omega_m$ in the modulated waveform [26]. For duty cycle of 20% [Fig. 2 (bottom row)], an observation of the in-phase signal clearly shows resonances up-to the third harmonic and the quadrature signal shows resonances up to fourth harmonic. Since the quadrature signal is sensitive to the signal phase, it allows detection of very weak signals at high magnetic fields.

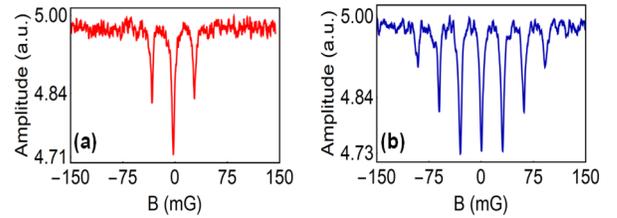

FIG. 3. Magneto-optical resonances: direct measurement using a low-pass filter for different modulation duty cycles of modulation (a) 50% and (b) 20% at $\Omega_m$ = 20 kHz.

Next, we measured the magnetic resonances directly using a low-pass filter for different duty cycles of light modulation [Fig. 3], in the same 90$^\circ$ fluorescence configuration. The cut-off frequency ($f_c$) for low-pass filter is set to 1 kHz and the signal is averaged over 100 samples using a digital oscilloscope. The low-pass filter measures only the signal amplitude i.e. insensitive to the signal phase. Signals measured with low-pass filter [Fig. 3] resemble the in-phase signals of lock-in amplifier [Fig. 2 (first column)]. The advantage of using low-pass filter is that the scanning rate of magnetic field (or modulation frequency at fixed field) can be faster compared to the lock-in detection where the scanning rate is limited by its long time-constant, hence it reduces the data processing time.

Figure 4 shows the plots of magneto-optical resonances when the magnetic field $B_y$ is fixed at 258.6 mG measured using a fluxgate sensor (Bartington, Mag-03) and the modulation frequency $\Omega_m$ of rectangular pulses of the laser beam is varied from 50 kHz to 200 kHz. Unlike lock-in detection, the low-pass filter is phase-insensitive and therefore, the resonance signal is measured over a wide range of $\Omega_m$ sweep. Laser beam modulation with 50% duty cycle shows magnetic resonance at Larmor frequency [Fig. 4(a)]. When the duty cycle is lowered to 20%, resonances also occur at the second and third sub-harmonics of Larmor frequency [Fig. 4(b)].

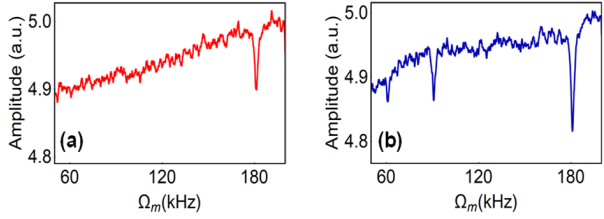

FIG. 4. Magneto-optical resonances observed by changing the laser beam modulation frequency $\Omega_m$: Direct measurement using low-pass filter at different duty cycles (a) 50% and (b) 20%.

The presence of cycling transition $F_g = 2 \rightarrow F_e = 3$ in Na $D_2$ line in a Doppler-broaden medium, is expected to increase the fluorescence (higher absorption) at the center of magnetic resonance [3], which is the characteristic of a bright resonance. However, contrary to this, we observed a dark resonance, i.e. a decrease in fluorescence at the center of resonance. To understand the observed behavior, we performed experiments with and without repump beam by switching the EOM on and off, respectively while locking the laser to $D_{2a}$ or $D_{2b}$ transition. These results are shown in Fig. 5. The resonance, obtained when the laser beam is tuned to $D_{2a}$ transition (without the repump beam), is shown in Fig. 5 (a). Average laser intensity in this case, is set to 0.25mW/cm$^2$ with 20% duty cycle and the magnetic field $B_y$ is kept fixed at 258.6 mG. Compared to the amplitude observed near the crossover, we observed a reduction in the amplitude of resonance by a factor of ~1.2. However, the sign of resonance did not change. When the EOM is switched on, one of the side bands produced by EOM becomes resonant with the Na $D_{2b}$ transition and acts as a repump beam. Considering a 50% loss of power in the two sidebands due to EOM, the total laser beam power is adjusted to bring the probe beam power to 0.25 mW/cm$^2$. This results in an increase of resonance width from 2.4 kHz to 3.7 kHz as measured from the fittings of experimental data [Fig 5 (top row red line)], and the signal amplitude is also increased now by a factor of ~1.8.

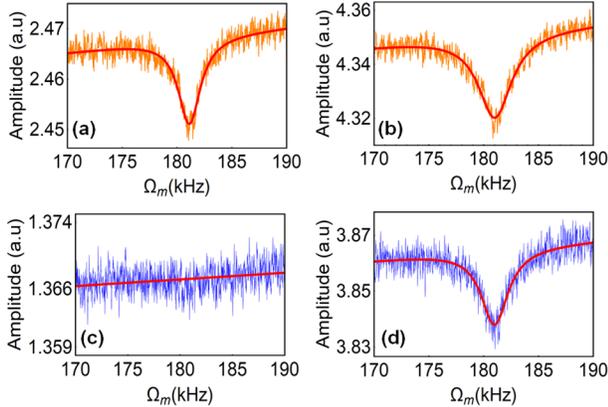

FIG. 5. Magneto-optical resonances without (first column) and with (second column) repump beam when the probe beam is locked to $D_{2a}$ (top row) and $D_{2b}$ (bottom row) transitions. The red line in each plot shows Lorentzian fit on the data.

Next, the laser is tuned to Na $D_{2b}$ transition [Fig. 5(c)] keeping all the other experimental parameters and conditions same as that for Fig. 5(a). Due to the presence of cycling transition $F_g = 1 \rightarrow F_e = 0$ in the Na $D_{2b}$ line, the resonance is expected to be a dark resonance. Experimentally, no resonance is observed in the signal [Fig. 5(c)] when the EOM is switched off. A dark resonance is observed only when the repump beam, resonant to $D_{2a}$ transition, is switched on [Fig. 5(d)]. In this case, we believe that the roles of probe and repump beams are reversed, which indicates that the magnetic dark resonance is only produced due to light absorption in the $D_{2a}$ transition. The $D_{2b}$ transition has less absorption compared to $D_{2a}$ transition [27], which suggests that magnetic resonance in $D_{2b}$ transition is weak and may not be possible to observe, due to low sodium density in the cell, under the present experimental conditions. We have confirmed this independently by observing a dark resonance in $D_{2b}$ transition in the sodium cell heated above 100° Celsius.

The observed dark resonance in $D_{2a}$ transition can be explained by considering excited state decoherence of $F_g = 2 \rightarrow F_e = 3$ transition due to collision of Na atoms with buffer gas atoms. The collisional decay typically suppresses the coherence transfer due to spontaneous emission from the excited state, which results in the formation of bright resonance for $F_g \rightarrow F_e = F_g+1$ transition [20, 28]. Therefore, the sign of resonance in Na $D_2$ line is reversed in the presence of buffer gas atoms. In the next section, we present a theoretical model by including the excited state decoherence to simulate the reversal from peak to dip of the magnetic resonance.

Remote magnetometry with mesospheric sodium atoms can only be performed with backscattered fluorescence [4,5]. Therefore, measuring magnetic resonances in backscattered fluorescence is consistent with remote magnetometry. Typically, in laboratory experiments, reflection from the cell window can prevent one to perform fluorescence measurements in the backward direction. In our experiment, an Na cell with a wedge window is used to eliminate backward reflection. Only half of the backscattered fluorescence is collected in reflection by the p-BS (shown in Fig. 1). Figure 6 shows magneto-optical resonances obtained using backscattered fluorescence light with the probe beam only (i.e. EOM is off). Light modulation frequency is set to 20 kHz at duty cycle 20% and the average intensity is set to 0.25 mW/cm$^2$. Figure 6(a) shows quadrature phase signal obtained by demodulation using the lock-in amplifier as B$_y$ scan is performed. The quadrature component shows resonances up to fourth harmonic, which matches our earlier observation in Fig. 2 (bottom row). However, as the in-phase component did not show any resonance due to weak backscattered fluorescence. We used a low-pass filter at $f_c$ =1 kHz to measure the resonance amplitude [Fig. 6(b)]. A faster magnetic field scan at 4 Hz allowed us to perform long sample averaging on the oscilloscope, and thereby, detecting the weak signal. Presence of three harmonics in the signal can be seen in Fig. 6(b). We also measured the resonances in backscattering by scanning the modulation frequency $\Omega_m$, as shown in Figs. 6(c) and(d). In Fig. 6(c), the measurements are performed at low magnetic field (B$_y$ = 29 mG) to observe resonances at subharmonics of the Larmor frequency. Although these resonances are weak, still the second and the third subharmonics are visible along with the first harmonic at Larmor frequency. Magnetic field is then increased to 258.6 mG [Fig. 6 (d)], which is comparable to earth's magnetic field. A resonance of width 2.6 kHz (3.7 mG) is observed at Larmor frequency corresponding to the applied magnetic field.

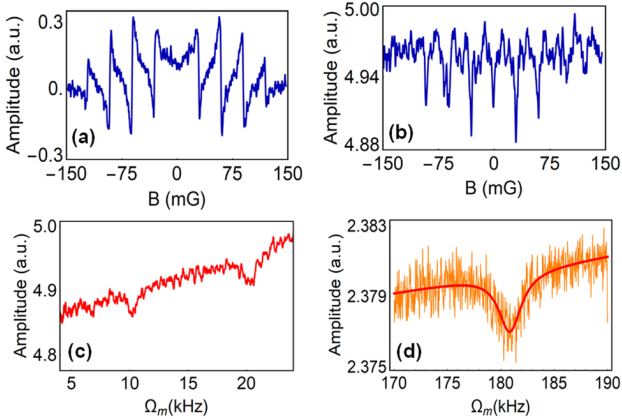

FIG. 6. Magneto-optical resonances obtained from backscattered fluorescence light using lock-in amplifier (a) and low-pass filter (b)-(d).

## IV. THEORETICAL ANALYSIS

As discussed in the experimental section above, we observed a dark magnetic resonance with a dip at the center, and we explained the origin of this dip being formed due to excited state decoherence caused by collision with buffer gas atoms. To further validate our observation, we have performed theoretical simulation of the magneto-optical resonance for $F_g = 2 \rightarrow F_e = 3$ transition using a model based on the density-matrix equations [29]. We consider that the electric field amplitude of light is modulated as

$$\vec{E}(t) = E_o(1 + \Delta M \cos[\Omega_m t])e^{-i\omega t}\hat{e}_L + c.c \qquad (1)$$

where $\omega$ is probe beam frequency, $\hat{e}_L$ defines the polarization vector, and $\Delta M$ is the depth of modulation. To match with our experimental configuration, the light polarization is considered to be circular with propagation direction along the z-axis. A fixed magnetic field B$_y$ is applied along the y-axis, perpendicular to the propagation direction z.

After making the rotating wave approximation to the light-atom interaction Hamiltonian, the evolution of atomic density matrix $\rho$ can be described by time-dependent Liouville equation:

$$\frac{\partial \rho}{\partial t} = \frac{1}{i\hbar}\left[\tilde{H}, \rho\right] - \frac{1}{2}\{\hat{\Gamma}, \rho\} - \frac{\gamma_{coll}}{2}\{P_e, \rho\} + R \qquad (2)$$

Here $\tilde{H}$ represents the total Hamiltonian of the atomic system in the rotating wave frame and $P_e$ is the projector operator corresponding to the excited state sublevels. The Hamiltonian $\tilde{H}$ includes the internal atomic energy levels, light-atom interaction, and magnetic field-atom interaction. The relaxation matrix $\hat{\Gamma}$ includes the spontaneous decay rate $\Gamma$ of the excited state and the transient relaxation rate $\gamma$ of each sublevel due to the exit of atoms from the laser beam. The effect of buffer gas collision is included in third term of eq. (2) by considering a dephasing rate $\gamma_{coll}$ of Zeeman coherence in the excited state sublevels [28]. Matrix R describes the repopulation of ground state sublevels due to decay rates $\Gamma$ and $\gamma$. The theoretical model is simplified by not considering the atomic motion, spatial distribution of light intensity, and the effect of neighboring transitions. To find the absorption spectra with amplitude-modulated light, we numerically solved the time-dependent density matrix equations. The absorption of the probe beam is calculated from the expression

$$\alpha \approx \sum_{e_i g_j} \frac{\Gamma}{\pi \Omega_R} \Big( \beta_{e_i g_j} \operatorname{Im}[\rho_{e_i g_j}(t)] + \beta'_{e_i g_j} \operatorname{Re}[\rho_{e_i g_j}(t)] \Big) \qquad (3)$$

where $\Omega_R = \langle F_g \|D\| F_e \rangle E_0$ is the reduced Rabi frequency ($\hbar = 1$) of the laser field and $D$ is a dipole operator. Here, the coefficients $\beta_{e_i g_j}$ and $\beta'_{e_i g_j}$ are dependent on the ground state and excited state sublevels involved in the optical transitions. Fluorescence signal measured in our experiment is directly proportional to the absorption signal described in eq. (3). Using eq. (3), the in-phase and quadrature components of the lock-in detection are found by demodulating the numerical solution for α at the first-harmonic of the modulation frequency [29].

Figure 7 shows the calculated in-phase and quadrature phase signals of probe absorption, as a function of modulation frequency for fixed magnetic field $B_y$ = 258.6 mG. In the absence of buffer gas collision ($\gamma_{coll}$ = 0), a bright resonance with peak in absorptions signal (or fluorescence signal) at the center of in-phase resonance is obtained as expected for $F_g$→$F_e$ = $F_g$+1 transition [Fig. 7(a)]. The corresponding quadrature component in Fig. 7(b) shows a dispersive behavior of the phase. When dephasing among excited state sublevels due to buffer gas collision is included ($\gamma_{coll}$ ≠ 0), the Zeeman coherence among the excited state sublevels gets suppressed before its transfer to the ground state via the spontaneous emission process. This results in the reduction of absorption or dip at the center of resonance as seen in [Fig. 7(b)], which agrees with the experimental result shown in Fig. 5(a). Similarly, the quadrature component of absorption also shows an opposite dispersive behavior when $\gamma_{coll}$ ≠ 0 [Fig. 7(d)] compared to $\gamma_{coll}$ = 0 [Fig. 7(c)]. Contrary to the explanation in reference [21], our observations suggest that the observed dip in magnetic resonance is not dependent on the detector position, but rather formed in a buffer cell due to collision induced decoherence irrespective of the scattering direction

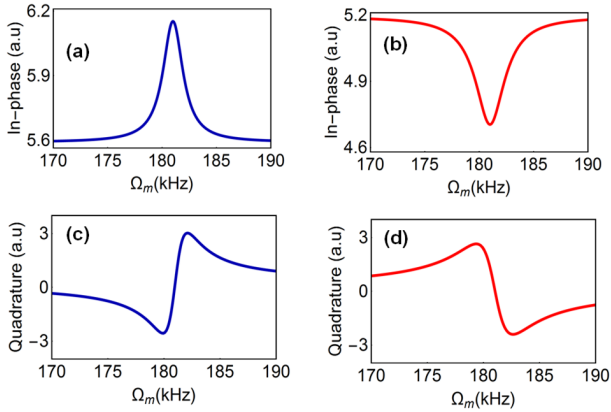

FIG. 7. Calculated magneto-optical resonances for $F_g$= 2→$F_e$= 3 transition: Collisional dephasing rate, $\gamma_{coll}$ = 0 (a) and (c), and $\gamma_{coll}$ = 1.5 Γ (b) and (d). Other parameters: Rabi frequency $\Omega_R$= 0.01 Γ and transit decay rate $\gamma$=$10^{-5}$ Γ. Magnetic field is set at 258.6 mG.

## IV. CONCLUSIONS

In conclusion, we have investigated magneto-optical resonances in sodium $D_2$ line using fluorescence from sodium atoms in a vapor cell containing Ne buffer gas. We performed measurements using fluorescence emitted both in perpendicular and backward direction to the laser beam propagation axis. Maximum amplitude of magnetic resonance was measured near the crossover peak. Due to collisional dephasing in the excited state of F=2 →F'=3 transition, a dark resonance is observed in sodium $D_2$ line. This is validated using a theoretical model. Resonance in $D_{2b}$ transition was also investigated in the presence of a repump beam. Higher-harmonic resonances and subharmonic resonances were observed with modulated light by employing respective magnetic field and modulation frequency scans. Resonances acquired using lock-in amplifier and low-pass filter were compared. Resonances measured in backscattered fluorescence were found considerably weaker but matched with the observed resonance in perpendicular direction. The present work has improved the understanding of magneto-optical resonances in sodium $D_2$ transitions, which will be useful in conducting remote magnetometry experiments in the near future.

## ACKNOWLEDGEMENTS

This work was supported by NASA EPSCoR funding #80NSSC17M0026, and NASA MIRO funding #NNX15AP84A.